\documentclass[letter]{ptptex}
\begin{document}
\markboth{
M. Kasai, H. Asada and T. Futamase
}{
}
\title{Toward a No-Go Theorem for an Accelerating Universe through a  
Nonlinear Backreaction
}
\author{
Masumi \textsc{Kasai},$^{1,}$\footnote{E-mail:
  kasai@phys.hirosaki-u.ac.jp} 
Hideki \textsc{Asada}$^{1,}$\footnote{E-mail:
  asada@phys.hirosaki-u.ac.jp} 
and 
Toshifumi \textsc{Futamase}$^{2,}$\footnote{E-mail:
  tof@astr.tohoku.ac.jp}
}
\inst{
$^1$Faculty of Science and Technology, Hirosaki University,
Hirosaki 036-8561, Japan\\
$^2$Astronomical Institute, Tohoku University, 
Sendai  980-8578, Japan
}
\notypesetlogo
\recdate{January 6, 2006}

\abst{
The backreaction of nonlinear inhomogeneities to the cosmic expansion 
is re-analyzed in the framework of general relativity. 
Apparent discrepancies regarding the effect of the nonlinear backreaction,
which exist among the results of previous works in different gauges, are
resolved. 
By defining the spatially averaged matter energy density 
as a conserved quantity in the large comoving volume, 
it is shown that the nonlinear backreaction 
neither accelerates nor decelerates the cosmic expansion 
in a matter-dominated universe. 
The present result in the Newtonian gauge is consistent with the
previous results obtained in the comoving synchronous gauge.  Although our
work does not give a complete proof, it strongly suggests the
following no-go theorem: No cosmic acceleration occurs as a result of
the 
nonlinear backreaction via averaging. 
}

\maketitle

The recent observation of the isotropy of the cosmic microwave background 
radiation (CMBR) \cite{WMAP} and large galaxy surveys,  
such as SDSS,\cite{SDSS} indicate 
that the universe is remarkably isotropic and homogeneous over scales 
larger than some 100 Mpc.  
However, it is not straightforward 
to describe the universe using an isotropic and homogeneous metric, namely, 
the Friedmann-Lemaitre-Robertson-Walker (FLRW) model, because the
local universe is in fact very inhomogeneous. 
The solution of 
the Einstein equation with an averaged homogeneous matter distribution 
is not a solution with a realistic  matter distribution,  
because of the nonlinearity of the Einstein equation.
Thus, it is naturally conjectured that the expansion law of the FLRW model 
may be modified by local inhomogeneities. In fact, there have been 
investigations studying this point,
\cite{Tomita88,Futamase89,Futamase96,Kasai92,Kasai93,Kasai95,Russ96,Russ97,SA,AK} 
and some modification has been reported,  
with apparent discrepancies among these works 
(for instance, see Refs.~\citen{Futamase96} and \citen{Russ97}). 
The result might depend on the choice of the coordinates as well as 
the definition of the averaging procedure. 
For these reasons, it has not been possible to clearly relate such a nonlinear
effect  with observations. 
Furthermore, recent observations of Type Ia supernovae 
\cite{Riess,Perlmutter} and the CMBR \cite{FGC,BC} 
strongly suggest that the cosmic expansion is accelerating. 
Understanding the source of this accelerated expansion is one of the 
greatest unsolved problems in modern cosmology \cite{Bahcall,PR}. 
This acceleration seems to require an unknown type of energy (dark energy, 
or perhaps a cosmological constant). 
A possible alternative idea to explain the acceleration is that 
the energy resulting inhomogeneities leads to additional terms 
in the Friedmann equation, as if dark energy existed. \cite{Rasanen,Kolb05} 
Before reaching a definite conclusion regarding effects due to 
the nonlinear backreaction, we must carefully elucidate the 
construction of a FLRW model with a mean density which obeys 
the equation of state (EOS) for dust matter. 
Otherwise, we would be misled to conclude that a
``correction''  
appears as a deviation from an averaged FLRW model, although 
the averaged matter energy density in such a model 
does not necessarily satisfy the EOS for dust matter. 

Here we attempt to clear up this confusion by carefully constructing 
an averaged FLRW model in the framework of general relativity. 
Because there is no unique choice of the averaged spacetime 
in the inhomogeneous universe, we take the point of view 
that the averaged density of the matter is conserved 
in a large comoving volume. Our choice seems natural and 
the most suitable for describing the averaged FLRW model 
with a mean density that satisfies the EOS for dust matter. 
The purpose of this Letter is to show that the nonlinear backreaction 
neither accelerates nor decelerates the cosmic expansion 
in a matter-dominated universe. It is also shown that 
the backreaction behaves like a small positive curvature term. 
We  show that the conclusion does not depend on 
the definition of the averaging procedure.

We restrict out study to the case of a vanishing shift vector, i.e., 
$N^i=0$. 
The line element is written as 
\begin{equation}
ds^2 = -(N dt)^2 + \gamma_{ij}dx^i dx^j . 
\end{equation}
The unit vector normal to a hypersurface foliated by $t =
\mbox{const}$  
is denoted by 
\begin{equation}
n^{\mu}=\left(\frac{1}{N}, 0, 0, 0\right) . 
\end{equation}
The extrinsic curvature is defined as  
\begin{equation}
K^i_{\ j} \equiv \frac{1}{2N}\gamma^{ik}\, \dot{\gamma}_{kj} , 
\end{equation}
where the dot denotes differentiation with respect to time. 
The Einstein equation is reduced to 
the Hamiltonian constraint, the momentum constraint and (the 
trace of) the evolution equation as \cite{AF}

\begin{eqnarray}
  {}^{\scriptscriptstyle(3)}\!R + \left(K^i_{\ i}\right)^2 
- K^i_{\ j} K^j_{\ i} &=& 16 \pi G E , \\
  K^j_{\ j|i} - K^j_{\ i|j} &=& 8\pi G J_i , \\
  \dot{K}^i_{\ i} + N K^i_{\ j} K^j_{\ i} - N^{|i}_{\ |i} 
&=& -4\pi G N(E+S) , 
\end{eqnarray}
where ${}^{\scriptscriptstyle(3)}\!R$ is the 3-dimensional Ricci 
scalar curvature, $|$
represents the 3-dimensional covariant derivative, and we have 
\begin{eqnarray}
  E &=& T_{\mu\nu} n^{\mu} n^{\nu} = \frac{1}{N^2} T_{00} ,\\
  J_i&=&-T_{\mu i} n^{\mu} = \frac{1}{N} T_{0i} , \\
  S &=& T_{ij} \gamma^{ij} .
\end{eqnarray}

The three-dimensional volume $V$ of a compact domain $D$ on a
$t=\mbox{const}$ hypersurface is
\begin{equation}
  V = \int_D \sqrt{\gamma}\, d^3x , 
\end{equation}
where 
\begin{equation}
\gamma = \det(\gamma_{ij}) . 
\end{equation}
Here, $V$ is considered to be a volume sufficiently large that we can 
assume periodic boundary conditions. 

The scale factor $a(t)$ is defined from the volume expansion rate of
the universe: \cite{Kasai92,Kasai93} 
\begin{equation}
  3 \frac{\dot{a}}{a} \equiv \frac{\dot{V}}{V} = \frac{1}{V}
  \int_D \frac{1}{2} \gamma^{ij}\,\dot{\gamma}_{kj}
  \sqrt{\gamma}\, d^3x .
\end{equation}
Next, we introduce the averaging procedure \cite{Kasai93}
\begin{equation}
  \langle{A}\rangle \equiv \frac{1}{V} \int_D A \sqrt{\gamma}\, d^3x . 
\end{equation}
With this, we find  
\begin{equation}
  3 \frac{\dot{a}}{a} = \langle{NK^i_{\ i}}\rangle . 
\end{equation}
Then we define $V^i_{\ j}$ as 
\begin{equation}
  V^i_{\ j} \equiv NK^i_{\ j} - \frac{\dot{a}}{a}\delta^i_{\ j} , 
\end{equation}
which represents the deviation from  uniform Hubble flow. 
With these definitions, one can show \(\langle{V^i_{\ i}}\rangle =0\). 

Averaging the Einstein equations, we obtain 
\begin{eqnarray}
  \left(\frac{\dot{a}}{a}\right)^2 &=& \frac{8\pi G}{3}\langle{N^2E}\rangle -
  \frac{1}{6} \langle{N^2\, {}^{\scriptscriptstyle(3)}\!R}\rangle -
  \frac{1}{6} \langle{(V^i_{\ i})^2 - V^i_{\ j} V^j_{\ i}}\rangle ,
  \label{eq:16}\\
  \frac{\ddot{a}}{a} &=&
  -\frac{4\pi G}{3}\langle{N^2(E+S)}\rangle
  + \frac{1}{3} \langle{(V^i_{\ i})^2 - V^i_{\ j} V^j_{\ i}}\rangle
  + \frac{1}{3} \langle{N N^{|i}_{\ |i} + \dot{N} K^i_{\ i}}\rangle . 
\label{eq:17}
\end{eqnarray}
We assume the matter of the universe to be irrotational dust. 
Then, the energy-momentum tensor is written  
\begin{equation}
T^{\mu\nu}=\rho u^{\mu}u^{\nu} , 
\end{equation}
where $u^{\mu}$ is the four velocity of the fluid flow. 

Up to this point, the treatment is fully general and exact. In order
to evaluate the R.H.S. of Eqs.~(\ref{eq:16}) and (\ref{eq:17}), we
need the actual metric of the inhomogeneous universe. 
We can proceed further either with an exact analytic approach employing a
class of exact inhomogeneous cosmological solutions or with an approximate
treatment of the inhomogeneous metric obtained  perturbatively. 
In the comoving synchronous
gauge, i.e., that with $N=1, u^{\mu}=(1,0,0,0)$, an exact analytical
treatment of 
Eqs.~(\ref{eq:16}) and (\ref{eq:17}) has been carried out by one of the
present authors\cite{Kasai92, Kasai93}.  
It has been shown that there exists a class of exact inhomogeneous
solutions that are, nevertheless, homogeneous and isotropic on
average, with no backreaction to the cosmic expansion.  
Approximate approaches in the form of perturbative analyses have also
been performed 
in several 
gauges, and some modifications have been reported with apparent
discrepancies among those works.\cite{Futamase96,Russ97}
In order to clear up this confusion, in the
following, we  re-examine the approximate approach and solve
Eqs.~(\ref{eq:16}) and (\ref{eq:17}) perturbatively by iteration. 

Let us start from the
Einstein-de Sitter background as the zeroth-order solution, for
simplicity. 
Next, we obtain the first-order inhomogeneous solution from the
linearized Einstein 
equation. 
It is sufficient to consider the cosmological post-Newtonian metric 
at linear order, \cite{Tomita88,Futamase89,Futamase96,SA}
\begin{equation}\label{eq:lis}
  ds^2 = -\left(1+2\phi(\mbox{\boldmath{$x$}})\right) dt^2
  + a^2 \left(1-2\phi(\mbox{\boldmath{$x$}})\right)\delta_{ij} \, 
  dx^i dx^j , 
\end{equation}
where $\delta_{ij}$ denotes the Kronecker delta. 
From the Einstein equation at  linear order, we obtain
\begin{eqnarray}\label{eq:20}
  \phi_{,i i}&=& \frac{3}{2}\dot{a}^2 \left(
    \frac{\rho-\rho_b}{\rho_b} + 2\phi\right) , \\
  \label{eq:21}
  v^i &\equiv& \frac{u^i}{u^0} = - \frac{2}{3 a \dot{a}}\phi_{,i} ,
\end{eqnarray}
where $\rho_b$ is the background density, and  contraction has been 
taken with $\delta_{ij}$. 
Note that Eq.~(\ref{eq:lis}), together with Eqs.~(\ref{eq:20}) and
(\ref{eq:21}), represents the first-order solution of the Einstein equation in
the Newtonian gauge. 

Once we obtain the first-order solution, we can perform the iteration
to solve Eqs.~(\ref{eq:16}) and 
(\ref{eq:17}) at the next order. Using the first-order solution to
the R.H.S. 
of  Eqs.~(\ref{eq:16}) and
(\ref{eq:17}), and retaining up to quadratic order in $\phi$, the averaged 
Einstein equation becomes 
\begin{equation}\label{eq:fr}
  \left(\frac{\dot{a}}{a}\right)^2 = \frac{8\pi G}{3}
  \langle{T_{00}}\rangle +
  \frac{1}{a^2} \langle{\phi_{,i}\phi_{,i}}\rangle ,
\end{equation}
\begin{equation}\label{eq:dda}
  \frac{\ddot{a}}{a} = -\frac{4\pi G}{3}
  \langle{T_{00} + \rho_b a^2 v^2}\rangle
  - \frac{1}{3a^2}\langle{\phi_{,i}\phi_{,i}}\rangle ,
\end{equation}
where $v^2=\delta_{ij}v^iv^j$. 

Equation~(\ref{eq:fr}) seems to indicate that the nonlinear backreaction
expressed by the second term on the R.H.S. 
might {\it increase} the expansion rate. 
However, this is not the case. 
In order to obtain the correct result, we have to determine the mean
energy density $\bar{\rho}$ of dust matter.  
Most previous works do not give careful consideration of this point. 
The mean density must satisfy
\begin{equation}\label{eq:dust}
  \dot{\bar{\rho}} + 3 \frac{\dot{a}}{a} \bar{\rho} =0 . 
\end{equation}
Without this consideration, the backreaction cannot be properly
related to 
a deviation from the Hubble expansion driven by the mean density.
Clearly, $\langle{T_{00}}\rangle$ does not act as $\bar{\rho}$. 

In order to guarantee that $\bar{\rho}$ satisfies Eq.~(\ref{eq:dust}), 
we have to define the mean density as 
\begin{equation}\label{eq:rhobar}
  \bar{\rho} \equiv 
\langle{T_{00}}\rangle + \rho_b a^2\langle{v^2}\rangle 
  +  \frac{1}{4\pi Ga^2} \langle{\phi_{,i}\phi_{,i}}\rangle
= \langle{T_{00}}\rangle + \frac{5}{12\pi Ga^2}
\langle{\phi_{,i}\phi_{,i}}\rangle, 
\end{equation}
where we have used Eq.~(\ref{eq:21}) to rewrite the term containing 
$\langle{v^2}\rangle$ as that with
$\langle{\phi_{,i}\phi_{,i}}\rangle$. 
The quantity $\bar{\rho}$ is uniquely determined in this form under
the following conditions: 
1) it is expressed as a linear combination of the terms appearing on the
R.H.S. of Eqs.~(\ref{eq:16}) and
(\ref{eq:17});  2) it  satisfies Eq.~(\ref{eq:dust}).
If these conditions are satisfied, then Eqs.~(\ref{eq:fr}) and (\ref{eq:dda})
can be rewritten as
\begin{equation}\label{eq:fr-2}
  \left(\frac{\dot{a}}{a}\right)^2 = \frac{8\pi G}{3}\bar{\rho} -
  \frac{1}{9a^2} \langle{\phi_{,i}\phi_{,i}}\rangle , 
\end{equation}
\begin{equation}\label{eq:dda-2}
  \frac{\ddot{a}}{a} = -\frac{4\pi G}{3}\bar{\rho} .  
\end{equation}
It is thus seen that we can safely call the second term on the R.H.S. 
of Eq.~(\ref{eq:fr-2}) the {\it nonlinear backreaction}.

Equations (\ref{eq:fr-2}) and (\ref{eq:dda-2}) reveal the following points. 
\begin{itemize}
\item The nonlinear backreaction neither accelerates nor decelerates 
the cosmic expansion. In other words, the cosmic acceleration 
$\ddot{a}/a$ is determined merely by the mean density. 
\item The nonlinear backreaction reduces the expansion rate
$\dot{a}/a$. 
\item The nonlinear backreaction is proportional to $a^{-2}$ 
in the averaged Friedmann equation. Hence, it behaves as 
a (small) positive curvature term in the Friedmann model. 
This correction might be measured in future space observations,  
such as a Planck mission. 
\end{itemize}

It should be noted that Eqs.~(\ref{eq:fr-2}) and (\ref{eq:dda-2}) 
do not rely on averaging procedures, though the explicit form of 
the mean density does. For instance, one may simply use 
$\langle A\rangle\equiv V^{-1}\int A d^3x$ \cite{Futamase89}. 
Changes induced by using a different averaging scheme occur 
simultaneously only inside $\bar\rho$ of 
Eqs.~(\ref{eq:fr-2}) and (\ref{eq:dda-2}). 
Furthermore, the nonlinear backreaction term is invariant  
to second order with respect to the choice of the averaging procedure. 

It is worthwhile  mentioning gauge issues in the 
backreaction problem. One may wonder what happens under a different 
gauge condition.  The comoving synchronous (CS) gauge has been employed to 
investigate the influence of the backreaction on quantities used to
determine the age of the universe 
\cite{Russ97} based on the relativistic version of 
a Zeldovich-type approximation.\cite{Kasai95,Russ96} 
Using the CS gauge, one can obtain equations similar to
Eqs.~(\ref{eq:fr-2}) and (\ref{eq:dda-2}). Hence, we reach the
conclusion that  there is no change in 
the cosmic acceleration, and the averaged Hubble equation has 
a positive spatial curvature as a correction.  
Therefore, it is strongly suggested that the above conclusion is valid 
for any choice of the  gauge. In practice, only the
post-Newtonian-type and CS gauges have been employed in previous studies of the nonlinear 
backreaction problem,  as far as the present authors are aware. 
With this in mind, our conclusion should be considered seriously. 

Finally, we make a few comments on  related works that have appeared
recently.  In \S 3 of Ref.~\citen{ishiwal}, Ishibashi and Wald
give an argument concerning the effect of the backreaction, employing
an averaging. In spite of their 
statement that they ``point out that our universe appears to be
described very accurately on all scales by a Newtonianly perturbed
FLRW metric'',  they restrict their analysis to the comoving
synchronous gauge. They only consider the ``smallness'' of 
possible nonlinear effects in the Newtonian gauge. 
With regard to this point, this
Letter gives a new result, which shows explicitly that no cosmic
acceleration occurs as a result of the nonlinear backreaction of a Newtonianly
perturbed FLRW metric.  Hirata and Seljak\cite{hirasel} also reached a
negative conclusion concerning the accelerating expansion due to superhorizon
cosmological perturbations. Their argument, however, does not treat
nonlinear effects due to local (subhorizon) 
inhomogeneities through averaging, as does the analysis carried out in
this Letter. 

\section*{Acknowledgements}
The authors would like to thank T. Kataoka for the detailed 
calculations in his master's thesis (Hirosaki Univ.).

\end{document}